\documentclass[12pts]{elsart3p}
\usepackage{graphicx}
\usepackage{mathpazo}
\usepackage{amsmath}
\newcommand{\comment}[1]{}
\usepackage{amssymb}  
\usepackage{color} 
\begin{document}
\begin{frontmatter}
\title{Reconstruction methods for acoustic particle detection in the deep sea using clusters of hydrophones}
\author[a1]{C. Richardt \corauthref{cor1}}
\ead{carsten.richardt@physik.uni-erlangen.de}
\corauth[cor1]{C. Richardt}
\author[a1]{G. Anton} 
\author[a1]{K. Graf}
\author[a1]{J. H\"o\ss l}
\author[a2]{A. Kappes}
\author[a1]{U. Katz}
\author[a1]{R. Lahmann}
\author[a3]{Ch. Naumann}
\author[a1]{M. Neff}
\author[a1]{F. Sch\"ock}
\address[a1]{Universit\"at Erlangen
ECAP (Erlangen Centre for Astroparticle Physics),
Erwin Rommel Str. 1,
91058 Erlangen, 
Germany}
\address[a2]{On leave of absence at the University of Wisconsin Madison, USA}
\address[a3]{CEA Saclay, IRFU 91191 Gif-sur-Yvette, France}

\begin{abstract}
This article focuses on techniques for acoustic noise reduction, signal filters and source reconstruction.
For noise reduction, bandpass filters and cross correlations are found to be efficient 
and fast ways to improve the signal to noise ratio and identify a possible neutrino-induced acoustic signal.
The reconstruction of the position of an acoustic point source in the sea is performed by using small-volume clusters of 
hydrophones ($\approx$1m$^3$) for direction reconstruction by a beamforming algorithm.  
The directional information from a number of such clusters  allows for position reconstruction. 
The algorithms for data filtering, direction and position reconstruction are explained and demonstrated using 
simulated data.
\end{abstract}

\begin{keyword}
acoustic particle detection, UHE neutrinos, neutrinos, signal processing filtering, beamforming
\end{keyword}
\end{frontmatter}

\section{Introduction}
\begin{figure}[ht]
  \centering
  \includegraphics[angle=0,width=0.49\textwidth]{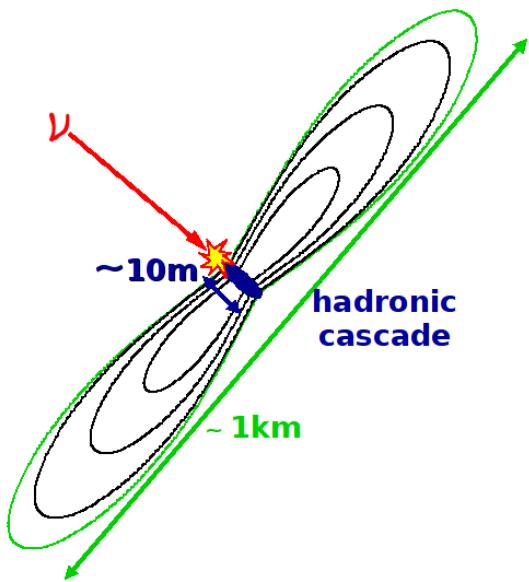}
  \caption {\label{fig:sw} {Neutrino-induced sound wave.  The typical shower length is 
  about $10$ m, resulting in a disc-shaped volume, in which the particle-induced signal can be detected.}}
\end{figure}

According to Greisen, Zatsepin and Kuzmin \cite{Greisen:1966jv,Zatsepin:1966jv} the flux of cosmic-ray protons 
observed on Earth should drop significantly at proton energies exceeding some $10^{19} \mbox{ eV }$ 
due to proton interactions with the cosmic microwave background. This drop in the flux is also known as the GZK-cutoff. 
Recent results from the Pierre-Auger observatory support these predictions \cite{CR_Spec_Auger}. 
In addition to the direct measurement of the proton flux, the detection of secondary particles resulting from the GZK mechanism 
offers a complementary way to investigate this effect.  
Along with other particles, the GZK mechanism produces neutrinos via pion and neutron decays.
Due to the low flux of expected GZK neutrinos, of the order of $100\mbox{ km}^{-2}\mbox{year}^{-1}$ \cite{TimoDiss},
huge detectors have to be built.  Conventional neutrino detectors 
measure the Cherenkov light emitted by secondary particles produced in a neutrino interaction. 
Such an optical detector would comprise an extremely large number of 
sensors because the attenuation length ($\sim60$ m \cite{Attenuation}) of light in water or ice is small.
Currently alternative methods of detecting ultra-high energy neutrinos are being investigated \cite{arena08}.
One of these methods is acoustic particle detection, for which an upper limit is derived in \cite{gratta}. 
Intensive studies are currently performed at various places to explore the potential of the acoustic detection technique.
In a recent survey \cite{Thompson2008}, an overview of these experimental activities is given.
The thermo-acoustic model, first discussed by G.A. Askarian in 1957 \cite{Askarian:1957,Askarian:1962}, predicts the creation
of sound waves produced by an instantaneous energy deposition of a particle cascade in a
medium. The energy deposition results in a local rise of temperature causing the medium to expand (or contract, depending 
on the medium's properties) which in turn creates a sound wave, see Fig. \ref{fig:sw}. 

\begin{figure}[ht]
\begin{center}
  \includegraphics[angle=0,width=0.49\textwidth]{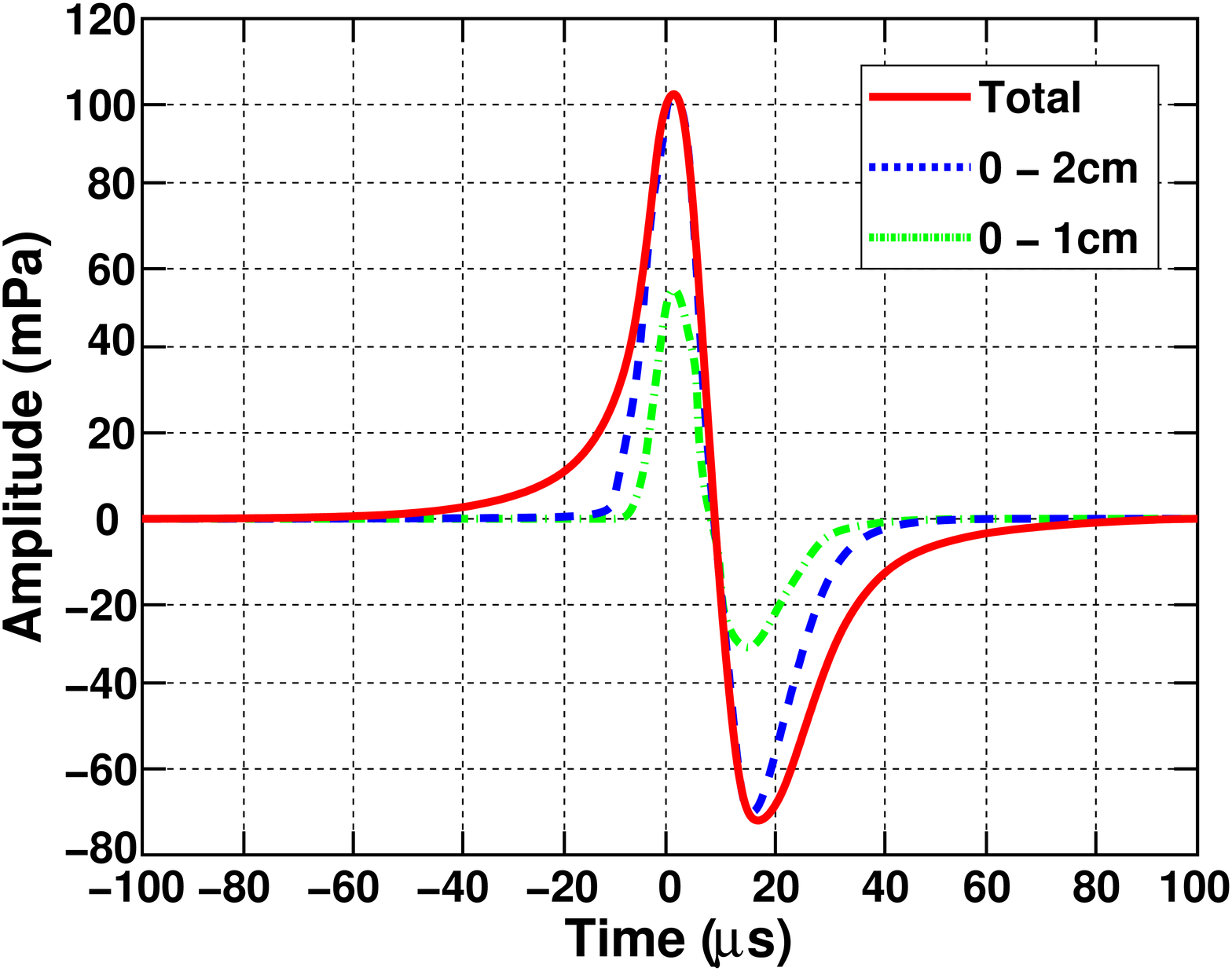}
  \caption {\label{fig:sig} {Simulation of a neutrino induced acoustic pulse for a cascade energy of 
  $100$ EeV at a distance of $1$ km from the source. The contributions of the energy deposits at distances from the 
    shower axis up to $1$ cm and $2$ cm and for the whole shower are shown.  Adapted from \cite{Sim_Acorne}.}}
\end{center}
\end{figure}

Test measurements verifying the model for water were performed using proton and laser beams \cite{Sulak:1978ay,Beam_Graf} 
and electron beams \cite{Bychkov}. 
The signal amplitude is found to depend approximately linearly on the temperature and vanishes at the maximum 
density thus demonstrating the thermo-acoustic origin of the signal.
An exemplary signal shape expected for a neutrino-induced hadronic shower is presented in Fig. \ref{fig:sig}.

The signal is bipolar, with a maximum of the power spectrum at about 15 kHz.
The attenuation length of an acoustic signal in water is of the order of $1000$ m at $20$ kHz \cite{TimoDiss},
yielding a large advantage over the optical method for the required large detector volumes.\\
In this article we describe techniques for data processing and source position reconstruction for acoustic particle detection. 
\section{Background and signal evaluation in the ocean}
\begin{figure}[ht]
\begin{center}
  \includegraphics[angle=0,width=0.49\textwidth]{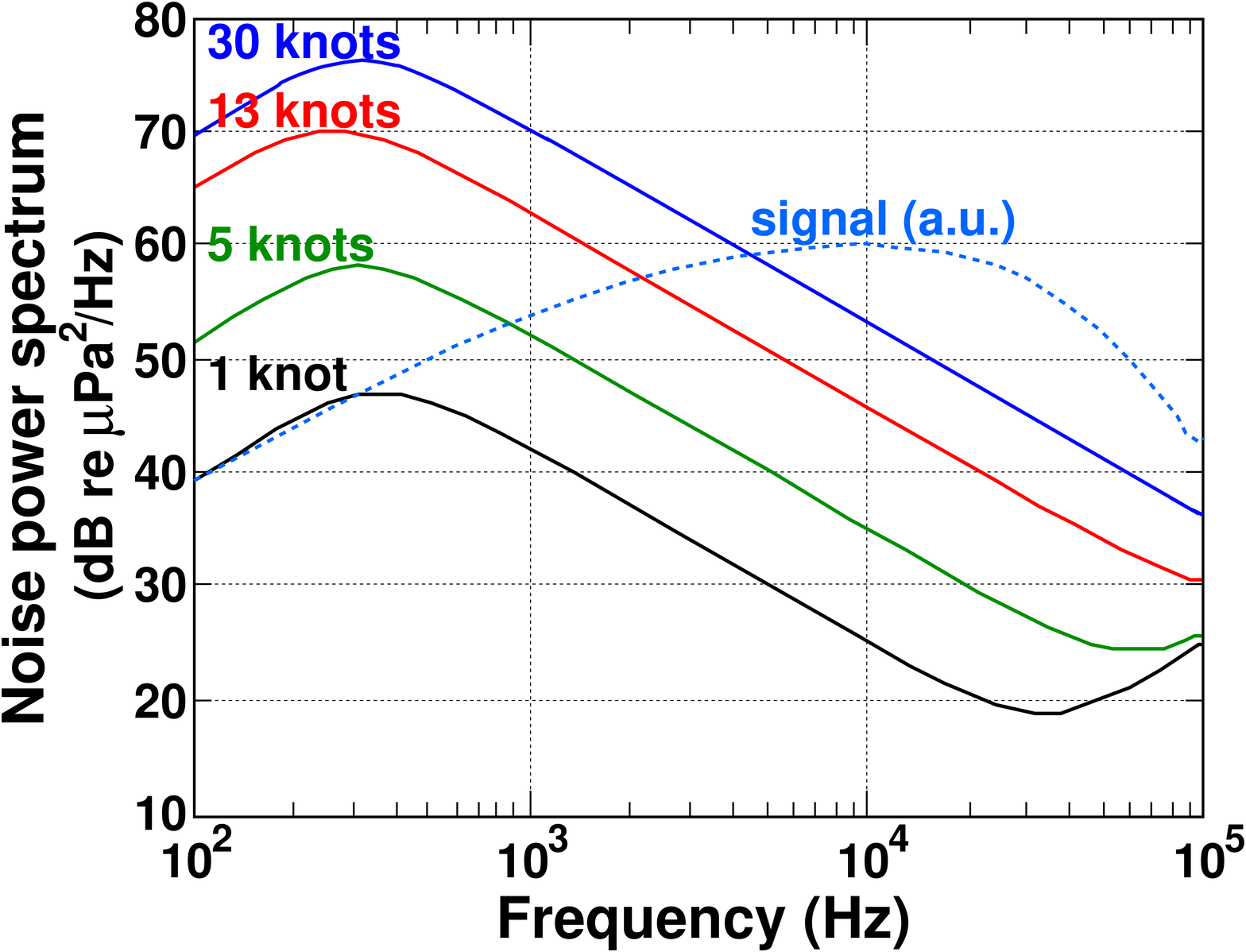}
  \caption {\label{fig:noisesp} {Power spectra from surface agitation and thermal noise for different wind speeds, 
     and the expected spectrum of the neutrino signal (arbitrary units)\cite{Lehtinen:2001km}.}}
\end{center}
\end{figure}

The challenge of acoustic particle detection is not only to develop sensors sensitive enough to detect 
neutrino-induced acoustic signals.  One essential problem is to separate the signal from the background 
from a wide range of sources, amongst them pressure changes due to current-induced turbulences, 
anthropogenic sounds (e.g. ships), wind and precipitation, biological sources (e.g. dolphins) and thermal noise. 
Figure \ref{fig:noisesp} shows the power spectrum of the surface-induced noise measured in the deep sea compared to the 
signal expectation. While the lower frequencies are dominated 
by wind and waves, the higher frequencies are dominated by the thermal noise \cite{Lehtinen:2001km}.

In the interesting frequency band for neutrino detection (about $5-50$ kHz), the wind-induced 
noise decreases with rising frequency, reaching a local minimum where the thermal noise starts to dominate.
The maximum in the frequency spectrum of a neutrino signal is close to the minimum of the noise spectrum.  
Integration of the frequency spectrum in the relevant frequency band results in a noise level of $10 - 100$ mPa. 
The expected signal amplitude is around the same value, for signals of a $10^{18}$eV shower at distances of about $100$ m.  
Therefore methods for noise reduction and signal filters have to be developed to enhance the particle detection sensitivity; 
examples are given in \cite{arena08,Lehtinen:2001km}.
Besides increasing the detection sensitivity one also has to understand the signal contamination resulting from 
point sources other than neutrinos. Of special interest are sources emitting neutrino-like, bipolar signals.
A prerequisite to ultimately calculate the sensitivity of an acoustic particle detector is thus the knowledge of the 
density of sources producing bipolar pulses, and the rate of such background signals.

\section{Data reduction}
\begin{figure}[ht]
\begin{center}
	\includegraphics[angle=0,width=0.49\textwidth]{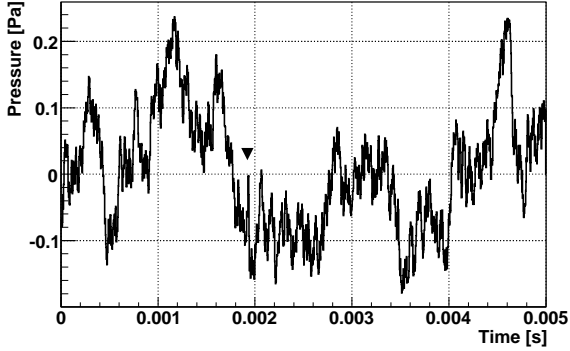}
     \caption {\label{fig:ex1} {Output of a single hydrophone with a noise level corresponding to $13$ knots of wind 
		and a signal of a $10^{19}$ eV shower at $200$m distance, occurring at about $0.0019$ s (see marker). }}
\end{center}
\end{figure}

\begin{figure}[ht]
\begin{center}
	\includegraphics[angle=0,width=0.49\textwidth]{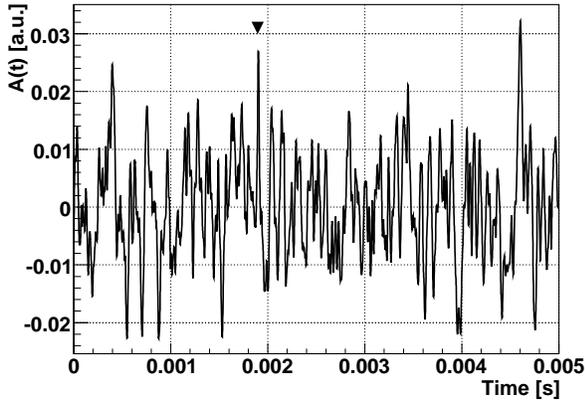}
     \caption {\label{fig:ex2} {After application of a cross correlation the lower frequencies are removed.  
	The signal to noise ratio is improved.  The marker indicates the signal position.}}
\end{center}
\end{figure}

\begin{figure}[ht]
\begin{center}
	\includegraphics[angle=0,width=0.49\textwidth]{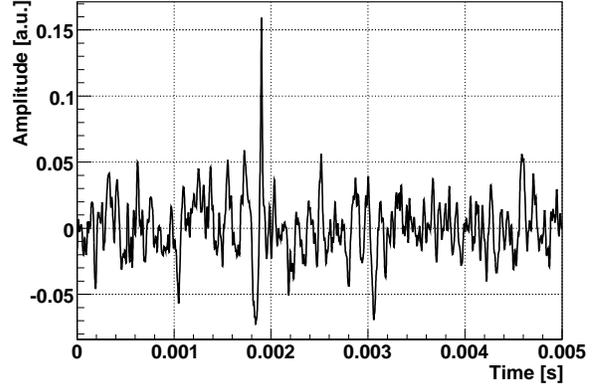}
     \caption {\label{fig:ex3} {Stacked output of six hydrophones. The signal at $0.0019$ s becomes
                clearly visible.}}
\end{center}
\end{figure}

Since neutrino-induced acoustic signals are primarily expected in a frequency band of 
$5-50\mbox{ kHz}$ the application of a bandpass filter significantly 
improves the signal to noise ratio. Simulations of neutrino-induced cascades with 
energies of $E_{\mbox{cascade}}>10^{19}$eV show that a band pass filter alone 
is sufficient to suppress the background to a level enabling the trigger of a possible signal, 
e.g. by a simple threshold trigger.  
Sensitivity for lower energies can be achieved by using the cross correlation\footnote{A technique very similar to 
matched filtering, described in e.g. \cite{arena06danaher}.}: 
\begin{equation}
  \begin{split}
	A(t)=([r + s]\otimes s)(t)= \\ \int_{-\tau/2}^{+\tau/2}d\tau^{'} [r + s](t+\tau^{'})s(\tau^{'})
  \end{split}
\end{equation}
where $s(\tau)$ is the expected signal template and $(r + s)(\tau)$ is the measured pressure amplitude 
containing background $r$ and possibly a signal $s$. $A(t)$ will have local maxima at times $t$ where the 
signal template matches the recorded data indicating the existence of a signal at a time $t$.
Another effective method to further improve the signal to noise ratio is {\it{stacking}}. Applied e.g. in geophysics 
for decades, this method uses arrays of sensors to identify a coherent signal. 
In the case of uncorrelated noise $r$ and a fully correlated signal $s$ in sensors $i,j$ the mean 
instantaneous signal power is
\begin{equation}
  \begin{split}
    \left<S^2\right>=\left < \left( \sum s_n\right )^2 \right >=N^2 \left< s^2\right >\\
    \mbox{    with }\left < s_is_j\right >= \left< s^2\right >\mbox{,}\nonumber
  \end{split}
\end{equation}
while the mean noise reduces to
\begin{equation}
  \begin{split}
    \left<R^2\right>=\left < \left( \sum r_n\right )^2 \right >=N \left< r^2\right >\\ 
    \mbox{    with }\left < r_ir_j\right >=\left< r^2\right >\delta_{ij}\mbox{,}\nonumber
  \end{split}
\end{equation}
thus resulting in a signal to noise ratio of
\begin{equation}
\frac{S}{R}= \sqrt{N}\frac{s}{r}
\label{eq:stack}
\end{equation}
where the sums run over the individual signals from all $N$ hydrophones and $\left< \mbox{ }\right >$ represents the mean. 
The $\sqrt N$-dependence is the best-case scenario for completely uncorrelated noise, but clearly indicates 
the benefit of using a larger number of sensors operated in clusters.
In geophysics this method can be applied passively by reading out the sensors in parallel since the 
duration of the signal is large compared to the difference in arrival time at the different sensors.
In our case an active procedure known as {\it{beamforming}} has to be used.  This method will be discussed in detail 
in section \ref{sec:dir}.
In Figs. \ref{fig:ex1} to \ref{fig:ex3} we demonstrate the application of the introduced methods for signal to noise 
ratio improvement. The data for the noise and the signal were simulated assuming a sampling frequency of $200$ kHz. 
The noise spectrum was modeled using the data  from Fig. \ref{fig:noisesp} for $13$ 
knots\footnote{$1$ knot $=1.852\mbox{ km/h}$} of wind. A cascade energy of $10^{19}$eV at a distance of $200$ m and six 
hydrophones have been assumed. For this example the signal arrival time is the same for all hydrophones. 
Figure \ref{fig:ex1} shows the output of a single 
hydrophone with a signal included at about $0.0019$ s. Identifying the signal is impossible in this case. 
Applying a cross correlation significantly improves the signal to noise ratio of a single hydrophone 
as can be seen in Fig. \ref{fig:ex2}. Stacking the output of all six hydrophones improves the signal to noise 
ratio to a point where a threshold trigger can be applied, see Fig. \ref{fig:ex3}.
Evidently a combination of all introduced methods is useful in order to successfully identify a signal. 
The advantage of using clusters of hydrophones in order to identify signals is a promising approach for 
future detectors.  Furthermore, a cluster greatly simplifies the 
reconstruction of sources, which will be discussed in the next section.

\section{Direction reconstruction}
\label{sec:dir}
\begin{figure}[ht]
\begin{center}
		\includegraphics[angle=0,width=0.49\textwidth]{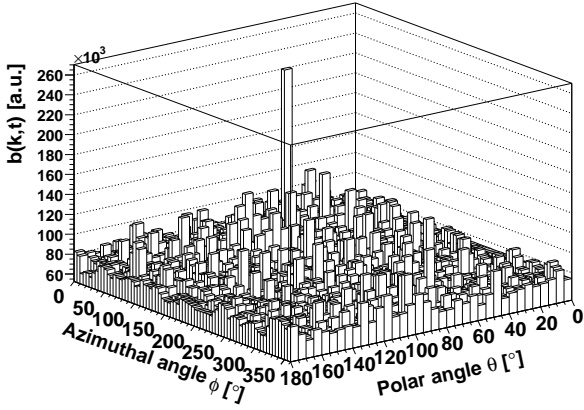}
		\caption {\label{fig:beamout1} {Beam-forming output for a set of simulated data including 
		   noise and a signal. The plot shows a clear peak for the direction of the signal. 
		The antenna configuration is the same as in Fig. \protect{\ref{fig:hydpos}}.}}
\end{center}
\end{figure}

Reconstructing the position of a point source by a small cluster of hydro\-phones 
($\approx 1\mbox{ m}^3$) becomes increasingly difficult the further the point source is away from
the cluster. However, direction reconstruction is well possible and therefore yields the prime observable for a small cluster. 
It is performed by creating a sound intensity plot scanning 
all directions in space ($4\pi$).  Given an array of $N$ hydrophones with 
coordinates $r_n\mbox{  } (n=1,2,...,N)$, the signals $p_n$ of every single hydrophone 
will be shifted in time corresponding to the difference in path length of the sound wave to reach the respective hydrophone.
Hence every direction in space corresponds to a set of time differences $\Delta t_n$ in the data.  
For a direction $\vec{k}$, the overall signal at the time $t$ is given by
\begin{equation}
\label{eq:beamer}
	b(\vec{k},t)=\sum_{n=1}^N w_np_n(t-\Delta t_n(\vec{k}))\mbox{ , }
\end{equation}
where the $w_n$ represent weighting factors for the hydrophones.  These factors can be adjusted to match 
the directional sensitivity of the hydrophones. In the following calculations $w_n\equiv 1$ was used.
The time differences $\Delta t_n$ are computed assuming a plane wave
\footnote{For point sources this is not entirely true, but a sufficiently good approximation 
if the distances are large compared to the dimensions of the hydrophone antenna.}.
The algorithm scans $4\pi$ with a predefined step size by applying the calculated time differences 
to the data, assuming a constant speed of sound. 
Once all directions are scanned, the maximum value of the produced output indicates the direction of the signal.  
Figure \ref{fig:beamout1} shows a sample output of the beamforming algorithm.

\section{Influence of antenna configurations}
\begin{figure}[ht]
	\begin{center}
        \includegraphics[angle=270,width=0.49\textwidth]{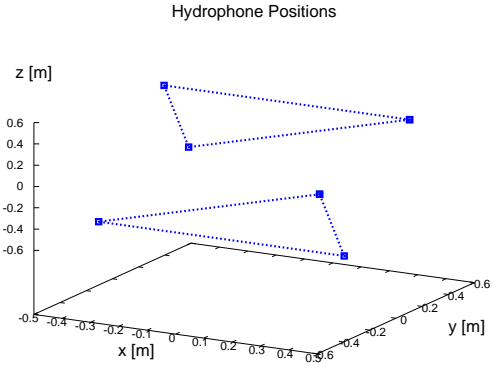}
        \caption {\label{fig:hydpos} {The hydrophone positions of one local cluster.}}
	\end{center}
\end{figure}

The time delays computed for the beamforming procedure depend
on the relative positions of the hydrophones in an antenna. Modifying
the geometrical arrangement of the hydrophones in an antenna will alter its directional
sensitivity. In this study different geometrical setups of six hydrophones were used (an example is shown in Fig. \ref{fig:hydpos}).

\begin{figure}[ht]
	\begin{center}
        \includegraphics[angle=0,width=0.49\textwidth]{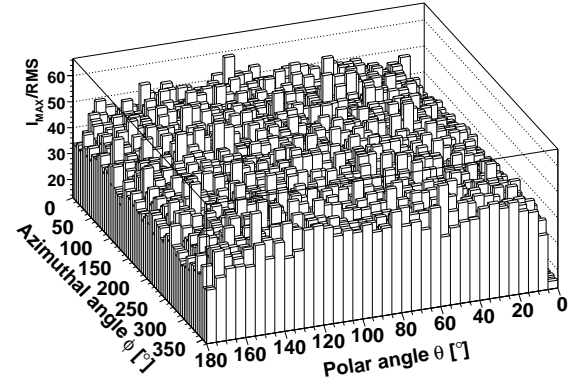}
        \caption {\label{fig:s1} {Maximum pressure amplitude from the beamforming algorithm 
	   divided by the RMS of the data, for all directions.}}
	\end{center}
\end{figure}

These geometries were evaluated with respect to their directional sensitivity 
by simulating arrival times of a wave front from all directions in steps of $5^\circ$.  
Noise samples were generated and signals added to the noise corresponding to the 
time delays for a given direction.
The maximum amplitude $I_{\mbox{max}}$ of the beamforming output divided by the RMS of the output was computed for all 
given directions. 
Figure \ref{fig:s1} shows a rather homogeneous reconstruction quality for all directions. 

The low $I_{\mbox{max}}$/RMS values for $\theta=0^{\circ}$ and
$\theta=180^{\circ}$ are a result of the symmetry of the detector in the given coordinate
system: If a signal comes directly from the top or the bottom of the antenna, all $\phi$
values for the beamforming algorithm are going to contribute with the same value.
Increasing or decreasing the antenna size by a factor of two creates a 
negligible difference.  As long as highly symmetric or two-dimensional geometries are avoided every setup 
will produce similar results.

\section{Source reconstruction}
\begin{figure}[ht]
	\begin{center}
        \includegraphics[angle=0,width=0.49\textwidth]{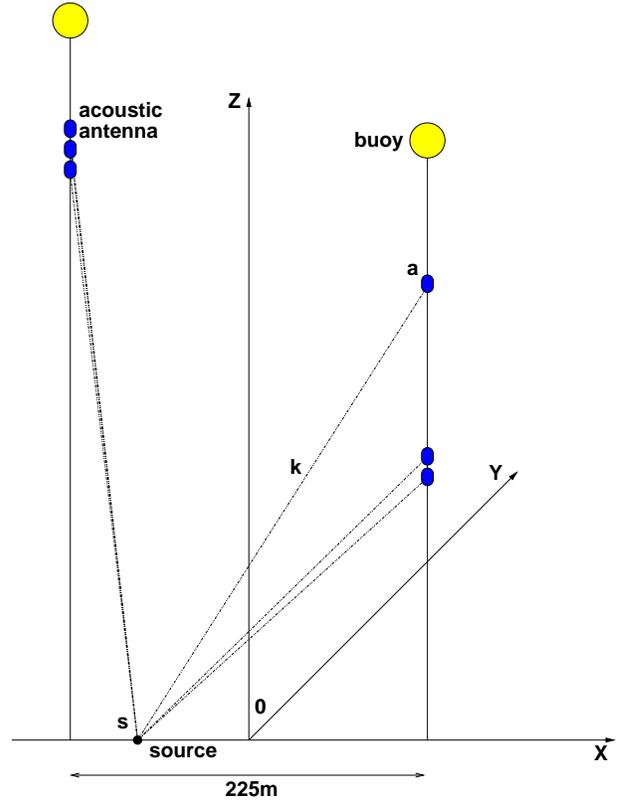}
        \caption {\label{fig:acousetup} {The geometry of the simulated acoustic setup.  The lines are $225$ m apart, 
	the vertical position of the stories are $180$ m , $195$ m and $305$ m for the line on the right 
	and $390$ m, $405$ m and $420$ m for the line on the left.}}
	\end{center}
\end{figure}

A simple technique for source position reconstruction using a given number of clusters is presented.
After the successful application of the beamforming algorithm each antenna $i$ with a detected signal 
will point to a direction $\vec{k_i}$ identifying where an event came from. 
Since the cluster positions $\vec{a_i}$ are known, one obtains a line pointing to the event for each cluster
\begin{equation}
	\vec{d_i}=\vec{a_i}+n_i\vec{k_i} \mbox{ , } n_i \in \mathbb{R}\mbox{.}
\end{equation}
The source position of a signal is in principle given by the intersection point of the lines $\vec{d_i}$. 
Due to uncertainties in the beamforming the reconstructed source location is where the lines are closest to each other.
Localising the point of closest approach for all lines is realized by calculating the square distance from a point 
$\vec{s}$ to the reconstructed lines 

\begin{equation}
\label{eqn:dist}
  \begin{split}
	L_i^2(\vec{s})=(\vec{s}-(\vec{a_i}+n_i\vec{k_i}))^2 \nonumber \\
	\mbox{    with } n_i=(\vec{s}-\vec{a_i})\vec{k_i}\nonumber \mbox{ , }
  \end{split}
\end{equation}

and minimising the sum of these distances.
This method was tested for a setup consisting of six antennas and was found to work well for distances up to $1$ km from the origin.
The test was conducted as follows:  First a source coordinate was defined, from where a  
spherical sound wave was emitted, propagating through the water at a constant speed of $1545\mbox{ ms}^{-1}$. 
As sketched in Fig. \ref{fig:sw}, a particle-induced sound pulse does not propagate as a spherical wave.  
However, since the technique discussed is independent of the radiation pattern a spherical wave was chosen as the most general case.  
A depth dependent velocity gradient was neglected because of the short distances evaluated with respect to the bending radius of 
about $90$ km \cite{TimoDiss}. The arrival times of the spherical wave at each single hydrophone were computed and the beamforming 
algorithm used to determine the directions of the wave onto the storey.  Those directions were then used to reconstruct 
the source location. See Fig. \ref{fig:acousetup} for the geometrical arrangement of acoustic sensors and the definition of the 
coordinate system. 

\begin{figure}[ht]
	\begin{center}
        \includegraphics[angle=0,width=0.49\textwidth]{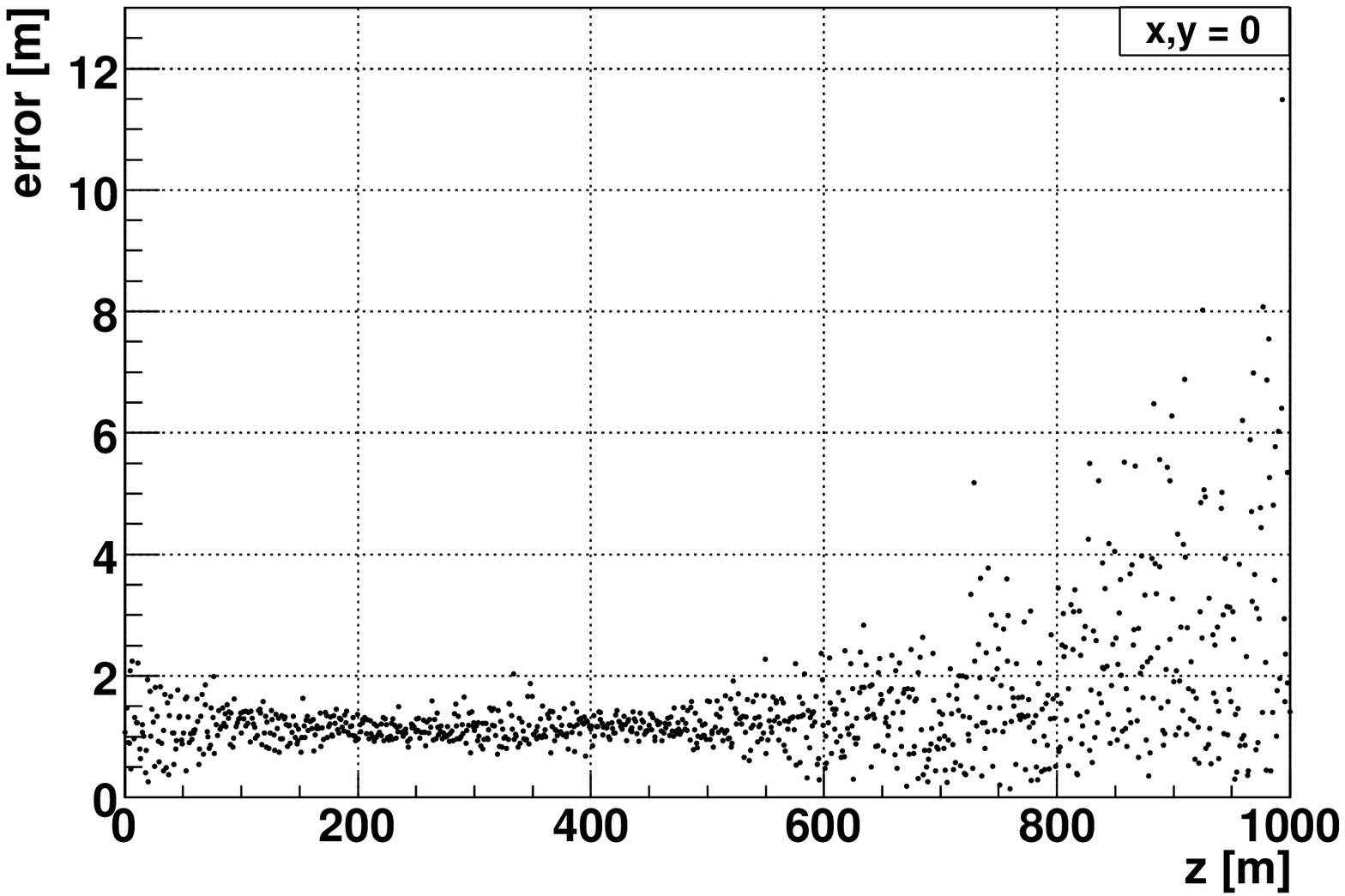}
        \includegraphics[angle=0,width=0.49\textwidth]{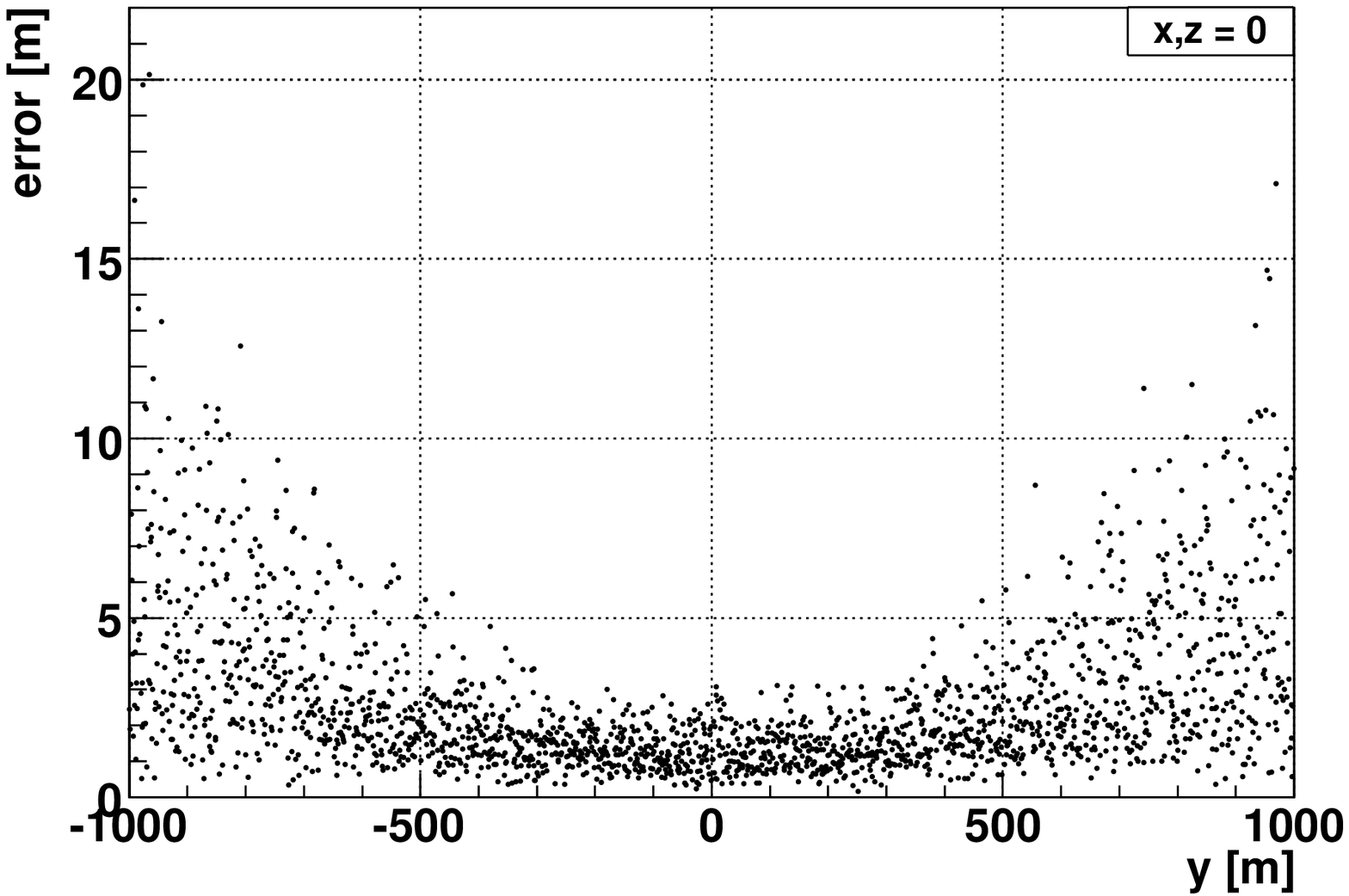}
        \includegraphics[angle=0,width=0.49\textwidth]{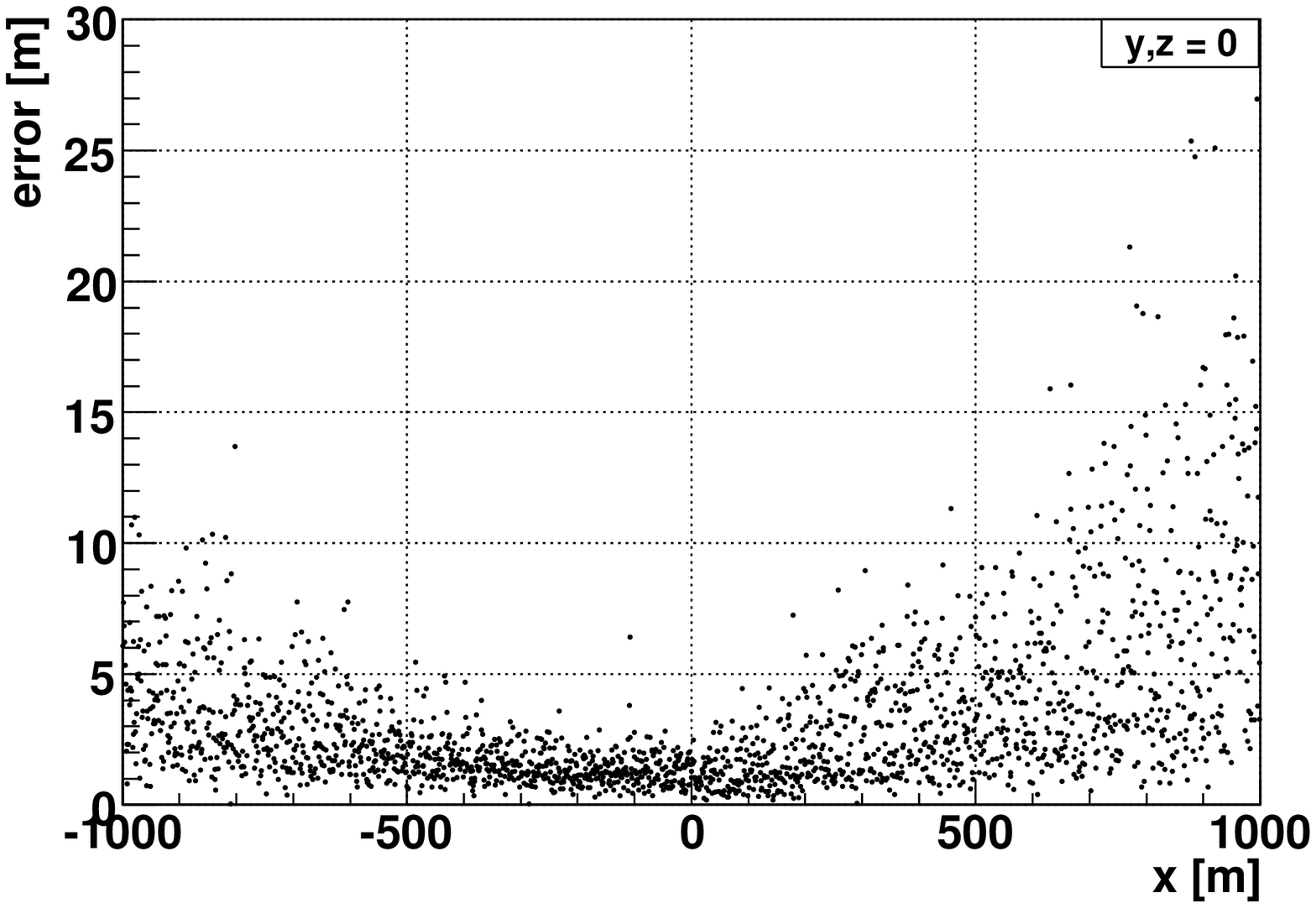}
        \caption {\label{fig:err} {The position reconstruction error for the cluster setup presented in Fig. 
	   \protect{\ref{fig:acousetup}}.
The error is the distance from the assumed source to the reconstructed source position. 
Top: Sources placed along the z-axis of the coordinate system.
Middle: Sources placed along the y-axis.
Bottom: Sources placed along th x-axis.}}
	\end{center}
\end{figure}

Typical errors were less than $10$ m for sources up to $500$ m away from the origin.
Figure \ref{fig:err} shows the dependence of the resulting error, i.e. the distance of the generated to the reconstructed source.
For each of the plots in Fig. \ref{fig:err} two of the source coordinates were kept constant while modifying the third.
The top plot in Fig. \ref{fig:err} for sources placed along the z-axis.  Up to $600$ m above the 
sea bed the error does not exceed $3$ m.  For z-coordinates above $600$ m the error increases due to the 
angular resolution (less than $0.5^{\circ}$) of the beamforming algorithm.  
The plot in the middle is for sources placed along the y-axis. The symmetry for the positive and 
negative y-values is due to the detector symmetry.  The error again increases with distance as a 
result of the angular resolution.  The bottom plot (sources placed along the x-axis) 
shows some asymmetry resulting from the fact that the clusters on the line on the left are 
higher than the clusters on the right resulting in best reconstruction values around the position of the line on the 
left. All plots demonstrate that using local clusters for direction reconstruction and using that information 
to reconstruct a source is a good approach for point source location.

\section{Summary and outlook}
This article focuses on techniques for acoustic noise reduction, signal filtering and source position reconstruction.
For noise reduction, bandpass filters and cross correlations prove to be efficient 
and fast means to improve the signal to noise ratio and to identify a possible neutrino-induced acoustic signal. 
The signal to noise ratio can be further improved by a factor of $\sqrt{n}$ ($n= $ number of hydrophones) by the use 
of local clusters and the application of stacking algorithms.  
Source reconstruction using directional information from localised clusters is a promising approach. 
Reconstruction errors for sources $250$ m from the center of the detector are of the order of a few meters, increasing for greater 
distances. The angular resolution is less than $0.5^{\circ}$.

\end{document}